\documentclass[pra, 12 pt]{revtex4}
\usepackage[english]{babel}
\usepackage{amsmath}
\usepackage{epsfig}

\begin{document}
%\draft
\title{Non-Planck equilibrium radiation in plasma model of early
Universe}
\author{S.A. Trigger}
\address{Joint\ Institute\, for\, High\,
Temperatures, Russian\, Academy\, of\, Sciences, 13/19, Izhorskaia
Str., Moscow\, 127412, Russia;\\
email:\,strig@gmx.net}

\begin{abstract}
Consideration of the adiabatic character of radiation expansion in
early Universe leads to the conclusion that equilibrium
distribution of the primordial radiation in the presence of
charged particles could be different from the Planck distribution
in some regions of the spectrum. The equilibrium distribution of
electromagnetic radiation (the black body radiation) is
generalized for the system containing an extremely dense fully
ionized plasma. The conditions of the adiabatic expansion of
radiation for the model of the early Universe are found.

PACS number(s): 42.50.Ar, 05.30.-d, 42.50.Ct,
52.25.-b \\

\end{abstract}.

\maketitle

The equilibrium distribution of photons in the background cosmic
radiation [1, 2] obeys the Planck law with high accuracy in the
microwave region of the spectrum. However, the primordial
radiation existed in the high density plasma medium. The plasma
composition in early Universe changed with time. Evolution of the
composition of the elementary particles in this plasma is not yet
clear enough. However, it is possible to consider the changes of
the temperature and density in the adiabatically expanding plasma
of a fixed composition.

The presence of the fully ionized plasma leads to the distortion
of the Planck distribution of radiation in low-frequency limit,
when the plasma frequency $\Omega_p$ is of the same order with the
current frequency. The thermodynamic functions of such radiation
are the functions of the dimensionless parameter
$\hbar\Omega_p/T$. When $\hbar\Omega_p/T\ll 1$, the thermodynamic
functions for radiation closely coincide with the ones obtained
with the Planck distribution. However, in the case when
$\hbar\Omega_p/T\geq 1$ the situation dramatically changes.
Realization of this latter condition in the early Universe seems
possible. In this case our understanding of the history of the
Universe is modified.

According to the Planck law [3, 4] the spectral energy
distribution of the black radiation in volume $V$ is equal
\begin{eqnarray}
dE(\omega) = \frac{V}{\pi^2 c^3}\frac{\hbar\omega^3
d\omega}{exp(\hbar\omega/T) -1}, \label{F1}
\end{eqnarray}
where c is the velocity of light. This formula implies that there
is equilibrium between photons and matter at temperature $T$.
Interaction between the photons and matter should be small enough
as not to disturb ideality of the photon gas. The net absorption
and creation of photons by matter in the volume $V$ are absent,
these processes are in a balance. Also, interaction between the
photons themselves is extremely weak. At the same time, to reach
equilibrium in the photon gas, a weak photon-matter interaction is
necessary. These conditions are satisfied quite well in gases for
all frequencies that lie far away from the frequencies of the
absorption lines of matter.

The Planck law has been verified experimentally, e.g. by pumping
laser radiation into the empty (or filled with a rarefied gas)
cavity through a small hole and then observing the spectral
distribution of the radiation emitted through the hole. The
spectral distribution of the emitted radiation is the Planck
distribution with high accuracy.

At the same time there exists a medium, plasma, in which the
electromagnetic radiation can exist without damping or with a weak
damping, and it possesses a different dispersion than that in
vacuum, where $\omega=ck$.

Let us consider a fully ionized plasma in some volume $V$. We
suppose that the plasma itself is in the equilibrium state, and
having the same temperature $T$ as the surrounding matter. The
condition of small damping of the transversal mode
$\omega=\sqrt{c^2 k^2+\Omega_p^2}$ in plasma [5] is
$\omega\gg\nu$, where $\Omega_p$ is the plasma frequency of the
electrons and $\nu$ is the characteristic collision frequency
between the charged particles. The characteristic values of the
collision frequency are equal to $\nu=\pi^{3/2}Z e^4 n
L/4\sqrt{2}\,T^{3/2}m^{1/2}$ for low frequencies, and to
$\nu_\omega=4\sqrt{2\pi}Ze^4nL/3T^{3/2}m^{1/2}$ - for high
frequencies. Here, $\nu\simeq\nu_\omega$ where $n$ is the electron
density and $L$ is the Coulomb logarithm for the electron-ion
collisions. Influence of the collisions on the radiation field is
small if $\Omega_p\gg \nu_\omega$ and $\omega\gg\nu_\omega$. The
condition of weak interaction between the particles can then be
written as
\begin{eqnarray}
\left(\frac{Z^2L}{4\pi}\right)^{1/3}\Gamma\ll1, \label{F2}
\end{eqnarray}
where we use the interaction parameter $\Gamma$ in the form
$\Gamma\equiv e^2 n^{1/3}/T$. It needs to be mentioned that for
one-component plasma the damping of oscillations is entirely
absent.

Below, we apply the general concept of stationary level of
quasiparticle excitations (here - electromagnetic waves) in
plasma: damping of electromagnetic waves is compensated for by the
permanent creation of the new waves in the equilibrium plasma at
constant temperature.

The number of oscillations in volume $V$ with the components of
the wave vector ${\bf k}$ in the intervals $d{\bf k}=dk_xdk_ydk_z$
is equal $V d^3 k/(2\pi)^3$. Then, the number of oscillations with
the modulus $k=\,\mid{\bf k}\mid$ equals $V k^2 dk/2\pi^2$. Taking
into account the existence of two independent polarizations of
oscillations and the dispersion equation for the transversal
oscillations in plasma $\omega=\sqrt{c^2 k^2+\Omega_p^2}$, we find
the number of oscillations with the frequency $\omega$ in the form
\begin{eqnarray}
\frac{V \sqrt{\omega^2-\Omega_p^2}\,\,\omega d\omega}{\pi^2c^3}.
\label{F3}
\end{eqnarray}
Naturally, the Bose distribution by frequencies is valid for the
transversal modes in plasmas, which are the Bose quasiparticles.
Therefore, the spectral distribution for electromagnetic waves in
plasma has the form
\begin{eqnarray}
dE (\omega)/d\omega=\frac{V \hbar\omega^2}{\pi^2c^3}\,
\frac{\sqrt{\omega^2-\Omega_p^2}\,}{exp (\hbar\omega/T)-1}\,
\theta(\omega-\Omega_p), \label{F4}
\end{eqnarray}
where $\theta(\omega-\Omega_p)$ denotes the unit-step function
which reflects the fact that transversal oscillations with
$\omega<\Omega_p$ are absent in plasma. By introducing the
dimensionless variable $q=\hbar\omega/T$ and the parameter
$a\equiv\hbar\Omega_p/T$, we can rewrite (\ref{F4}) in the form:
\begin{eqnarray}
dE (q)/dq=\frac{V T^4}{\hbar^3\pi^2c^3}\, \frac{q^2
\sqrt{q^2-a^2}\,}{exp (q)-1}\, \theta(q-a). \label{F5}
\end{eqnarray}
The changes in spectral curves as a function of $q$ for various
parameters $a$ are given in the Fig. 1.

In the plasma, the position of maximum in the spectral
distribution is a function of the parameter $a$ (Fig.~2).

For $a\rightarrow0$ the maximum value approaches the Planck limit
$q^0_{max}=2,822$. For the case $a\gg1$ the position of maximum
approaches $q^1_{max}\approx a + \frac{1}{2a}$ (according to
Eq.~(\ref{F5}). Now, we have $q_{max}>a$ for any $a$, and not only
for the asymptotically large values of the parameter $a$. In the
classical limit, we find $dE (\omega)/d\omega$ from Eq.~(\ref{F4})
\begin{eqnarray}
dE (\omega)/d\omega=\frac{V \omega T}{\pi^2c^3}\,
\sqrt{\omega^2-\Omega_p^2}\,\, \theta(\omega-\Omega_p). \label{F6}
\end{eqnarray}
Formula (\ref{F4}) is a generalization of the Rayleigh-Jeans
classical result (e.g. [6]) for the case of fully-ionized plasma.
Formally, Eq.~(\ref{F5}) can be easily obtained by calculating the
energy $\Xi$ of electromagnetic field in a transparent medium [7]
\begin{eqnarray}
\Xi=\frac{1}{4 \pi}\, \left\{(E^2)_\omega \frac{d[\omega
\varepsilon(\omega)]}{d \omega}+(H^2)_\omega \right\}, \label{F61}
\end{eqnarray}
if we use for the frequency-dependent dielectric function
$\varepsilon(\omega)$ in plasma (the case of a small damping, when
$\mbox{Im} \varepsilon(\omega)\rightarrow0$) the expression
$\varepsilon=1-\Omega_p^2/\omega^2$ and take into account the
expressions for the correlation functions of the electrical
$(E^2)_\omega$ and of the magnetic $(H^2)_\omega$ fields
\begin{eqnarray}
\varepsilon(\omega)(E^2)_\omega = (H^2)_\omega = \frac{2 \omega^3
\hbar\; \varepsilon^{3/2}(\omega)}{c^3} \coth
\left(\frac{\hbar\omega}{2T}\right). \label{F62}
\end{eqnarray}

It is necessary to stress here that for the ultra-relativistic
temperatures, when $T> m c^2$ (here $m$ is the mass of the charged
particles which determines the dominating contribution in the
current plasma frequency of the primordial plasma), the spectrum
of the transversal mode is different from $\omega=\sqrt
{\Omega_p^2+c^2k^2}$ and depends on the temperature. The spectrum
of the transversal mode is determined by the dielectric function
\begin{eqnarray}
\varepsilon^{tr}=1+\frac{\pi e^2 n c}{\omega k
T}\left\{-\frac{2\omega}{ck}+\left(1-\frac{\omega^2}{c^2
k^2}\right)\ln |\frac{ck-\omega}{ck+\omega}|\right\}, \label{F62a}
\end{eqnarray}
where the charge $e$ and the density $n$ are related to the
dominant (for determination of the transversal plasma mode)
fraction of the plasma, existing at any given moment. In general,
the expression for the spectrum cannot be written in analytical
form. For the region of frequencies $\omega^2\gg 2\pi e^2 n c^2/
T$, the radiation distribution is close to the Planck
distribution, because the spectrum approximately is described by
the formula $\omega^2 \simeq 2 \pi n e^2 c^2/T + c^2 k^2 \simeq
c^2 k^2$ . The asymptotic behavior of the spectrum for a small $k$
has the form
\begin{eqnarray}
\omega^2=\frac{4\pi e^2 n c^2}{T} + \frac{3}{2}\,c^2 k^2
\label{F62b}
\end{eqnarray}
Therefore, the gap in the spectrum of radiation distribution and
the deviation from the Planck distribution are structurally
similar to Eq.~(\ref{F4}). However, in the case of the
ultra-relativistic temperatures, the deviation appears near the
value of $\omega^2=\frac{4\pi e^2 n c^2}{T}\ll \Omega_p^2$,
instead of $\omega=\Omega_p$ being valid in the region of the
temperatures $T\ll mc^2$.

Let's now calculate the thermodynamic properties of the
equilibrium electromagnetic radiation in plasma (for simplicity we
restricted the consideration below by the region of the
non-relativistic temperatures, where the analysis can be done in a
more general form). Since the chemical potential is equal to zero,
the thermodynamic potential $\Omega$ coincides with the free
energy $F$, which can be written in the form
\begin{eqnarray}
F=\frac{8 \pi V T}{(2\pi)^3c^3}\, \int_{\Omega_p}^\infty d\omega
\omega \sqrt{\omega^2-\Omega_p^2}\,\ln[1-\exp(-\hbar\omega/T)].
\label{F7}
\end{eqnarray}
We suppose that the quasiparticles in plasma can be in first
approximation considered as an ideal gas. Then Eq.~(\ref{F7}) can
be rewritten as
\begin{eqnarray}
F=-\frac{V T^4}{3\pi^2 c^3 \hbar^3}\,
\int_{\hbar\Omega_p/T}^\infty dx [x^2-(\hbar\Omega_p/T)^2]^{3/2}
\,\frac{1}{\exp(x)-1}. \label{F8}
\end{eqnarray}
Therefore, the free energy can be represented as
\begin{eqnarray}
F=F_0 \phi(\hbar\Omega_p/T), \label{F9}
\end{eqnarray}
where
\begin{eqnarray}
\phi(\hbar\Omega_p/T)=\frac{15}{\pi^4}\,
\int_{\hbar\Omega_p/T}^\infty dx [x^2-(\hbar\Omega_p/T)^2]^{3/2}
\,\frac{1}{\exp(x)-1}. \label{F10}
\end{eqnarray}
The function $F_0=-V\pi^2T^4/45(\hbar c)^3$ is the free energy for
the free photon gas for the case when plasma is absent. For small
and large values of the ratio $\hbar\Omega_p/T$, the function
$\phi$ can be  written as
\begin{eqnarray}
\phi(\hbar\Omega_p/T\rightarrow0)\simeq 1-\frac{5}{2
\pi^2}\left(\hbar\Omega_p/T\right)^2, \label{F12}
\end{eqnarray}
and
\begin{eqnarray}
\phi(\hbar\Omega_p/T\gg1)\simeq \frac{45 \sqrt
3}{\pi^4}(\hbar\Omega_p/T)^{3/2} \,exp(-\hbar\Omega_p/T),
\label{F13}
\end{eqnarray}
respectively. The function $\phi(a\equiv\hbar\Omega_p/T)$ is shown
on Fig. 3.

The entropy of radiation $S=-\partial F / \partial T$ is expressed
as
\begin{eqnarray}
S=-\frac{4F_0}{T}\left[\phi(\hbar\Omega_p/T)+\frac{3(\hbar\Omega_p)^2}{4T^2}\chi(\hbar\Omega_p/T)\right],
\label{F14}
\end{eqnarray}
where the function $\chi$ is determined by
\begin{eqnarray}
\chi(\hbar\Omega_p/T)=\frac{15}{\pi^4}\,
\int_{\hbar\Omega_p/T}^\infty dx [x^2-(\hbar\Omega_p/T)^2]^{1/2}
\,\frac{1}{\exp(x)-1}, \label{F15}
\end{eqnarray}

Since $\chi(0)=5/2\pi^2$, the entropy for the case $a\ll1$ is
equal:
\begin{eqnarray}
S\simeq-\frac{4F_0}{T}\left[1-\frac{5a^2}{8\pi^2}\right].
\label{F16}
\end{eqnarray}
It means that for $a \ll 1$ the adiabatic expansion of the
radiation field is close to the case of the ideal photon gas
$TV^{\gamma_0-1}$, where $\gamma_0=4/3$. The small correction
$\gamma'$ to the value of $\gamma_0$ ($\gamma\equiv
\gamma_0+\gamma'$) is a function of the system's parameters
$\gamma'\simeq -5a^2/24\pi^2 ln V$.

For the opposite case, when $a \gg 1$,  taking into account that
$\chi(a \gg 1)=15 exp \,(-a)/\pi^4$, we find the entropy
\begin{eqnarray}
S\simeq-\frac{4F_0}{T}\frac{45 a^2 exp \,(-a)}{4 \pi^4}= \frac{V T
\Omega_p^2
 exp\, (-\hbar\Omega_p/T)}{\pi^2 \hbar c^3}. \label{F16}
\end{eqnarray}

Let us apply this theory to matter and radiation in the early
Universe at the stage when the matter was still in the plasma-like
state (primordial plasma), but the equilibrium between radiation
and plasma had already been reached. We will now determine the
time-dependent plasma frequency $\Omega_p(t)=\Sigma_\alpha
\Omega_p^\alpha (t)$, which is the sum of the different
time-dependent plasma frequencies of the charged particles
existing at the moment $t$. Time dependence is the consequence of
the expansion of the Universe in time $V=V(t)$ and, the connected
with this expansion, transformation of the number and species of
particles $N_\alpha(t)$ in primordial plasma. If one denotes, in
the model, the moment of the primordial plasma recombination as
$t_0$, and uses the adiabatic relation for radiation field
$V^{\gamma-1}T=const$ with $\gamma=4/3$ (which follows from the
conservation of the Planck radiation entropy $S_P$), then
\begin{eqnarray}
S_P=\frac{4F_0}{T}=\frac{4\pi^2 V T^3}{45 (\hbar c)^3}.
\label{F17}
\end{eqnarray}
In reality, the recombination process took time which is small in
comparison with the typical expansion time. Let us suggest, in
accordance with the model of Hot Universe, that in the moment
$t_0$, the Planck distribution for radiation was valid, as the
density was too low to satisfy the inequality
$\hbar\Omega_p(t_0)/T(t_0)\ll 1$. The accepted estimation for the
characteristic recombination temperature is $T(t_0)\sim 4000°K$.
Let us consider the behavior of parameter
$a(t)\equiv\hbar\Omega_p(t)/T(t)$ for earlier time $t \ll t_0$.
Taking into account Eq.~(\ref{F17}) we find
\begin{eqnarray}
a(t)=a(t_0)\left(\frac{V(t_0)}{V(t)}\right)^{1/6}=a(t_0)\left(\frac{T(t)}{T(t_0)}\right)^{1/2}.
\label{F18}
\end{eqnarray}
Therefore, for the model of the early Universe, the parameter
$a(t)$ decreases in time and for $t \ll t_0$ can be of the same
order or bigger than unity. For simplicity we suggest that the
quantity of the lightest ($\alpha=\alpha_0$) charged particles is
conserved, i.e. $N_{\alpha_0} (t) = const$. The more accurate
condition for the decrease of the parameter $a(t)$ in time, which
takes into account the changing composition of the charged
particles in primordial plasma, requires the increase in time of
the ratio $\Omega_p(t)V^{1/3}(t)/\Omega_p(t_0)V^{1/3}(t_0)$. The
condition $a(t)\geq 1$ is attainable if the period of time between
the set-up of the time-dependent equilibrium state for radiation
field in the primordial plasma and its recombination $\tau_0$, is
of the same order or bigger than $\tau=t_1-t_0$, where $t_1$ is
the root of the equation $a(t_1)=1$. If the condition $a(t)\gg 1$
is fulfilled, then, we have the isentropic process with the
entropy determined by Eq.~(\ref{F16}). In this case (if the number
of the light charged particles is conserved) we find the time
evolution of the parameter $a(t)$ for $t<t' \ll t_1$ (in the
region $a(t)\gg 1$) in the form
\begin{eqnarray}
a(t)=a(t')+ \ln \left(\frac{T(t)}{T(t')}\right). \label{F19}
\end{eqnarray}
Therefore, the parameter $a(t)$ decreases in time for the region
$a \ll 1$, as well as for the region where $a\gg 1$. If the state
of the plasma and the radiation field at the moment of
recombination are characterized by the parameter $a(t_0)\ll 1$ and
this state of the system is the result of the adiabatic expansion
of another state at the moment $t \ll t_0$ with the parameter
$a(t)>1$, we arrive at
\begin{eqnarray}
\frac{T(t_0)}{T(t)}=\frac{a^2(t_0)}{a^2(t)}\left[\phi(a(t))+\frac{3a^2(t)}{4}\chi(a(t))\right].
\label{F20}
\end{eqnarray}
The adiabat for the radiation field is shown on the Fig.4.

It is easy to see that the process of adiabatic expansion of the
primordial radiation with the parameter $a(t)> 1$ would lead to
the different future scenarios of the Universe evolution, when the
equilibrium radiation field is described after Planck, and by
using the generalized (for the case of arbitrary
$\hbar\Omega_p/T$) distributions.

\section*{Acknowledgment}

The author thanks to W. Ebeling, M.V. Fedorov, A.M. Ignatov, M.B.
Pyshnov, A.A. Rukhadze and P.P.J.M. Schram for valuable
discussions, and D. Montgomery who had attracted my interest to
the problem of equilibrium radiation in plasma at extremal
parameters. This work has been supported by The Netherlands
Organization for Scientific Research (NWO) and the Russian
Foundation for Basic Research.\\

% otsyuda ya dobavil figuri, podpravte podpisi sami

\begin{figure}[h]
    \epsfxsize=8.cm
    \epsfysize=8.cm
\epsfbox{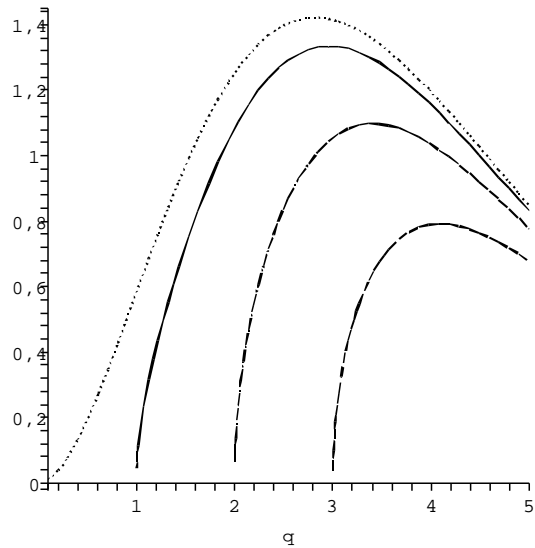} % from uni
  \caption{ The dimensionless spectral distributions dE(q)/dq for a=0
    (Planck distribution), and $a=1,2,3$ (non-Planck equilibrium radiation, the curves are from the top down)  as
    functions of the parameter $q=\hbar\omega/T$.
The parameter $a$ equals $a=\hbar\Omega_p/T$.}

    \label{fig1}
\end{figure}

    \begin{figure}[h]
    \epsfxsize=8.cm
    \epsfysize=8.cm
\epsfbox{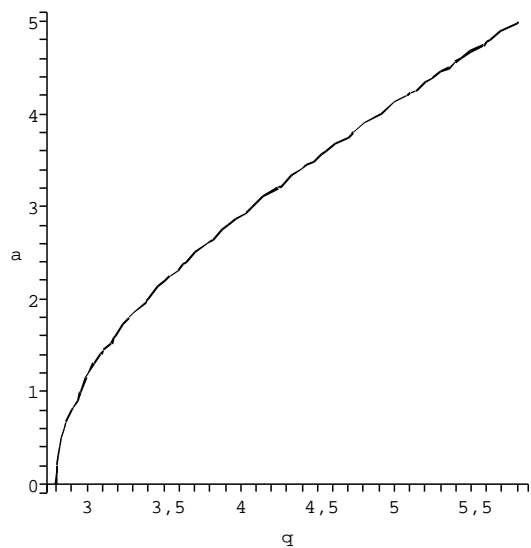} % from uni
  \caption{ The positions of maximums $q$ of the equilibrium radiation density in plasma as
function of the parameter $a$. }
    \label{fig2}
\end{figure}

\begin{figure}[h]
    \epsfxsize=8.cm
    \epsfysize=8.cm
\epsfbox{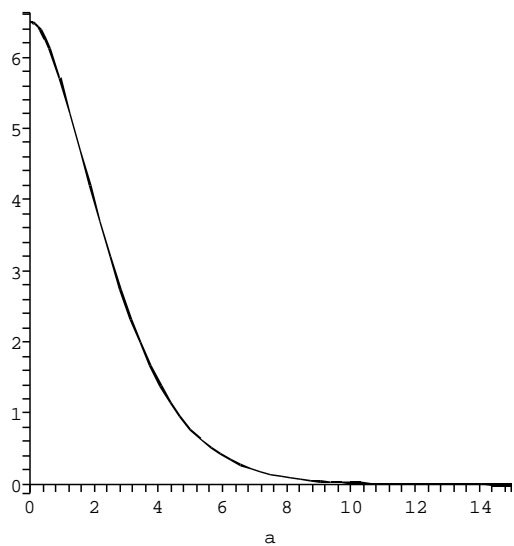} % from uni
  \caption{ The function $\phi$ determining the free energy $F=F_0\phi(a)$.}
    \label{fig3}
\end{figure}

\begin{figure}[h]
    \epsfxsize=8.cm
    \epsfysize=8.cm
\epsfbox{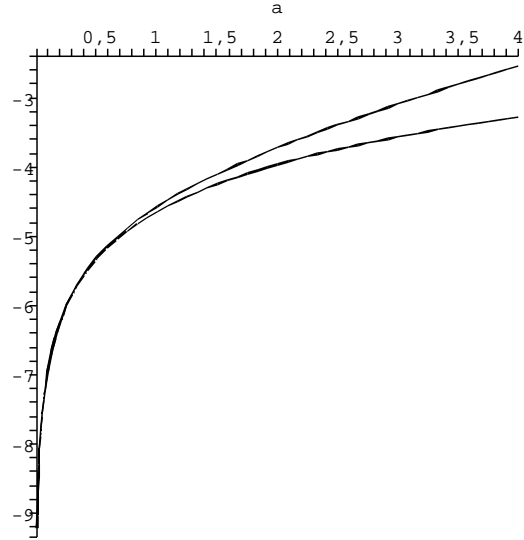} % from uni
  \caption{The ratio $a(t_0)$ as a function of the values of the parameter $a(t)$ for the fixed value of
  the parameter $T(t_0)/T(t)=0.0001$. The upper curve: the arbitrary values of the parameter
  $a$; calculation based on non-Planck distribution.
  The lower curve is calculated with the Planck distribution extended to the region $a>1$,
  where this distribution is not valid.
  The curves are plotted in the logarithmic scale. }

    \label{fig4}
\end{figure}

% konets moey dobavki

\end{document}